\documentclass[acmsmall,screen]{acmart}

\AtBeginDocument{%
  }

\setcopyright{acmlicensed}
\copyrightyear{2018}
\acmYear{2018}
\acmDOI{XXXXXXX.XXXXXXX}

\usepackage{multirow}
\acmJournal{JACM}
\acmVolume{37}
\acmNumber{4}
\acmArticle{111}
\acmMonth{8}
\usepackage{arydshln}

\begin{document}

\title{Quality Assessment in the Era of Large Models: A Survey}

\author{Zicheng Zhang}
\affiliation{%
  \institution{Shanghai Jiao Tong University}
  \city{Shanghai}
  \country{China}
}

\author{Yingjie Zhou}
\affiliation{%
  \institution{Shanghai Jiao Tong University}
  \city{Shanghai}
  \country{China}
}

\author{Chunyi Li}
\affiliation{%
  \institution{Shanghai Jiao Tong University}
  \city{Shanghai}
  \country{China}
}

\author{Baixuan Zhao}
\affiliation{%
  \institution{Shanghai Jiao Tong University}
  \city{Shanghai}
  \country{China}
}

\author{Xiaohong Liu$^\dagger$}
\affiliation{%
  \institution{Shanghai Jiao Tong University}
  \city{Shanghai}
  \country{China}
}

\author{Guangtao Zhai$^\dagger$}
\affiliation{%
  \institution{Shanghai Jiao Tong University}
  \city{Shanghai}
  \country{China}
}

\renewcommand{\shortauthors}{Zhang et al.}

\begin{abstract}
Quality assessment, which evaluates the visual quality level of multimedia experiences, has garnered significant attention from researchers and has evolved substantially through dedicated efforts. Before the advent of large models, quality assessment typically relied on small expert models tailored for specific tasks. While these smaller models are effective at handling their designated tasks and predicting quality levels, they often lack explainability and robustness. 
With the advancement of large models, which align more closely with human cognitive and perceptual processes, many researchers are now leveraging the prior knowledge embedded in these large models for quality assessment tasks.
This emergence of quality assessment within the context of large models motivates us to provide a comprehensive review focusing on two key aspects: 1) the assessment of large models, and 2) the role of large models in assessment tasks. We begin by reflecting on the historical development of quality assessment. Subsequently, we move to detailed discussions of related works concerning quality assessment in the era of large models. Finally, we offer insights into the future progression and potential pathways for quality assessment in this new era.
We hope this survey will enable a rapid understanding of the development of quality assessment in the era of large models and inspire further advancements in the field.
\end{abstract}

\ccsdesc[500]{Computing methodologies~Artificial intelligence}

\keywords{Quality assessment, large model, multimedia}

\received{20 February 2007}
\received[revised]{12 March 2009}
\received[accepted]{5 June 2009}

\maketitle

\section{Introduction}
Thanks to the rising demand for high-quality multimedia consumption and an enhanced Quality of Experience (QoE), specialized quality assessment tools have been developed to predict the quality levels of various media types. These tools are instrumental in improving a wide range of applications, including low-level enhancements (such as dehazing, brightening, and denoising), as well as compression/transmission systems, and 3D scanning and reconstruction processes. Quality assessment serves as a guiding light, directing the optimization of numerous algorithms, which is crucial and fundamental for both academic and industry. Specifically, quality assessment can be divided into subjective quality assessment and objective quality. Subjective assessment is often regarded as the most reliable and accurate method because the human visual system (HVS) is the ultimate recipient of visual signals in most visual communication systems. However, subjective testing is time-consuming and costly, and it cannot be directly integrated as an optimization metric in practical systems. Objective quality assessment methods, which are typically designed or trained using data from subjective assessments, can automatically predict visual quality. These methods are ideal for timely evaluation and optimization of system performance.

In the era before large models, quality assessment typically focuses on specific domains such as image quality assessment (IQA), aesthetic quality assessment (AQA), video quality assessment (VQA), and 3D quality assessment (3DQA). IQA deals with traditional technical distortions like noise, blur, and compression artifacts. AQA evaluates the artistic and compositional elements of images, assessing factors such as color harmony, balance, and emotional impact. VQA addresses temporal issues in video streams, including frame rate, resolution, and motion artifacts, to ensure smooth and clear playback. Meanwhile, 3DQA focuses on the visual quality of 3D models, including point clouds, voxels, and meshes, evaluating aspects like texture, resolution, and structural accuracy. These specialized assessments help optimize multimedia content across various platforms and devices. Many researchers employ small expert models based on handcrafted features with support vector regression (SVR), convolution neural networks, or transformers to deal with quality assessment tasks. These expert models are typically trained and validated using databases specific to particular domains. However, they often lack the ability to adapt directly to other domains and provide scores without accompanying explanations.

\begin{figure}
    \centering
    \includegraphics[width=.9\linewidth]{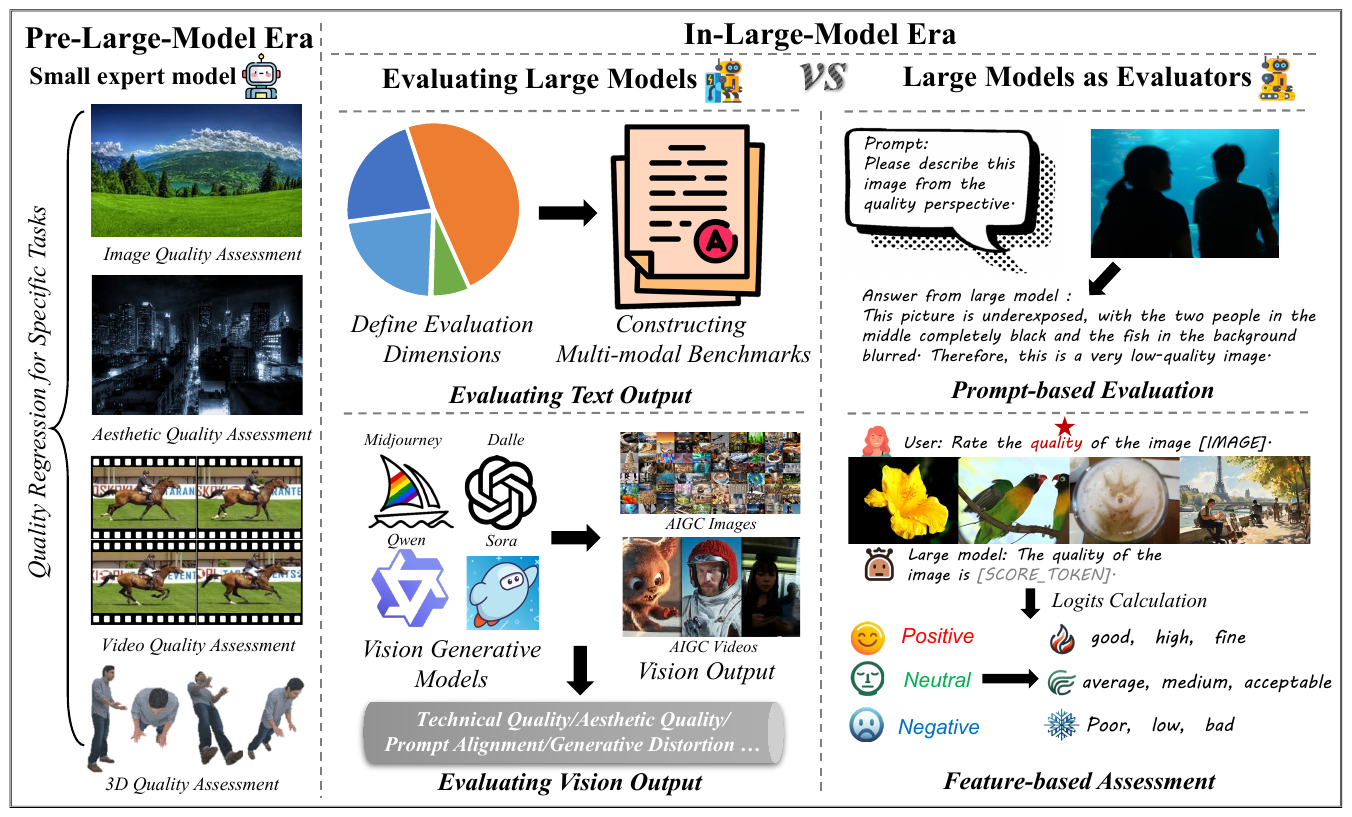}
    \caption{Illustration of the quality assessment development of pre-large-model era and in-large-model era.}
    \label{fig:framework}
    \vspace{-15pt}
\end{figure}

In the era of large models, we can leverage large language models (LLMs) and large multi-modal models (LMMs) to provide detailed descriptions of quality information and answer questions related to the quality of multimedia content. This approach enhances the explainability of quality assessments. Additionally, the rise of AI-generated content (AIGC) necessitates robust quality assessment tools to monitor and enhance generation quality. Consequently, numerous new quality assessment benchmarks and methods have been developed to evaluate the multimedia outputs of these large models. For instance, various multi-modal benchmarks have been introduced to assess the text response quality of LMMs, and many quality assessment methods are now focusing on predicting quality in a unified manner, rather than being confined to specific domains. Considering the significant changes in the quality assessment field, there is a pressing need for a comprehensive and updated survey. This survey would provide a better top-down understanding of the history, current state-of-the-art, and future trends in the field. Our survey is constructed on the core question:

\textit{How to evaluate large models and how to utilize large models as evaluators?}

As limited by pages, we confine our survey pool to encompass only papers that are considered important in the corresponding field. Specifically as shown in Fig. \ref{fig:framework}, our survey is organized as follows:
\begin{enumerate}
\item We begin by reflecting on the development of quality assessment before the advent of large models, offering a comprehensive comparison with the developments during the large-model era.
\item We summarize the advancements in large multi-modal models (LMMs) and the corresponding multi-modal benchmarks. Additionally, we explore the emerging focus areas and the developmental psychology behind quality assessment for AI-generated content (AIGC).
\item We review the methods of using large models as evaluators, discussing the differences and relationships among various techniques.  
\end{enumerate}

\section{Quality Assessment in the Pre-Large-Model Era}
In this section, we explore the landscape of quality assessment before the integration of large models. We will examine the methodologies (Image/Aesthetic/Video/3D Quality) as well as classic datasets (shown in Table~\ref{tab:qa}) traditionally employed, the limitations of these earlier approaches, and how they laid the groundwork for the development and necessity of large model frameworks.

\begin{table}[]
    \centering
    \caption{Brief comparison of the quality assessment databases in the Pre-Large-Model Era.}
    \resizebox{\linewidth}{!}{\begin{tabular}{l|c|c|c|c}
         \toprule
         Dataset & Year & Scale & Content &Task \\ \hline
         \multicolumn{5}{l}{\textit{Quality assessment for \textbf{Images}}}\\ \hdashline
         LIVE \cite{sheikh2005live}  & 2004  &779 & Compressed, blurred, and noisy images&Traditional image quality assessment\\
         TID2008 \cite{ponomarenko2009tid2008}& 2008 & 1,700 & 17 different types of distorted images  &Full Reference Visual Quality Assessment\\
         TID2013 \cite{ponomarenko2015image}& 2013 & 3,000 & 24 different types of distorted images  &Full Reference Visual Quality Assessment  \\
         KADID-10K \cite{kadid}&2019 &10,125 & 24 different types of distorted images  &Weakly-supervised image quality assessment\\
         
         MDID2013 \cite{gu2014hybrid} &2013 &324 & Compressed, blurred, and noisy images & Image quality assessment for multiple distortions\\
         MDID2016 \cite{sun2017mdid} &2016 &1,600 & Images subject to random types and degrees of distortion&Image quality assessment for multiple distortions\\
         DHQ \cite{min2018objective} &2019 &1750 &Dehazed images of real scenes &Dehazed image quality assessment\\
         SHRQ \cite{min2019quality} &2019 &600 & Synthetic hazy images &Dehazed image quality assessment\\
         DIBR \cite{bosc2011towards}&2011 &84 &Depth image-based rendering & Synthetic view quality assessment\\    
         SIQAD \cite{yang2014subjective} &2014 &980 &7 Commonly distorted screen content images &Screen image quality assessment\\
         SCIQ \cite{ni2017esim} &2017 &1800 & 9 Commonly distorted screen content images &Screen image quality assessment\\
         CCT \cite{min2017unified} &2019 &1320 &Natural scene, computer-generated, screen content images &Unified blind image quality assessment\\
         KonIQ-10k \cite{koniq} &2018 & 10,073& Images from public multimedia dataset YFCC100m \cite{yfcc} &Quality Assessment of authentically distorted Images\\
         CGIQA \cite{zhang2023subjective}&2023 &6000 &Games, movies with in-the-wild distortion& Quality assessment for computer graphics images\\ \hline
         
         \multicolumn{5}{l}{\textit{Quality assessment for \textbf{Aesthetic}}}\\ \hdashline
         AVA \cite{avaiaa} &2012 &255,530 & Images and metadata from the web&Improve performance on preference tasks\\ 
         CUHK-PQ \cite{luo2011content} &2011 &17,613 &7 different types of photo content & Content-based photo quality assessment\\
         AADB \cite{aadb} &2016 &10,000 &Photographic images on the web & Ranking the aesthetics of images\\
         PCCD \cite{chang2017aesthetic} &2017 &4,235 &Images from professional photographers &Assess photo aesthetics and photographic techniques\\
         DPC-Captions \cite{jin2019aesthetic} &2019 &154,384 & Images labeled with 5 aesthetic attributes&Aesthetic attributes assessment of images\\
         TAD66K \cite{he2022rethinking} &2022 &66,327&Images of 47 themes &Theme-oriented image aesthetic quality assessment\\
         ICAA17K \cite{he2023thinking} &2023 &17,726 &Images covering 30 color combinations &Image color aesthetics assessment\\
         SPAQ \cite{spaq} & 2020 & 11,125 & Images taken by 66 smartphones & Perceived quality of smartphone photography \\
         PARA \cite{piaadataset} &2022 &31,220& Images with wealthy annotations&Personalized image aesthetics assessment\\  \hline
         \multicolumn{5}{l}{\textit{Quality assessment for \textbf{Videos}}}\\ \hdashline
         LIVE-VQA \cite{livevqa} &2008 &160 &Videos of natural scenes with 4 types of distortions&Full-reference video quality assessment\\
         CSIQ \cite{csiqvqa} & 2014 &228 &Videos with 6 types of distortions &Full-reference video quality assessment\\ 
         CVD2014 \cite{cvd} & 2014 &234 &234 videos recorded using 78 different cameras &Quality assessment of video captured by cameras\\
         LIVE-Qualcomm \cite{qualcomm} &2016 &208 & Videos with 6 common in-capture distortions& Mobile in-capture video quality assessment\\
         KoNViD-1K \cite{kv1k} &2017 &1200 &Public-domain video sequences from YFCC100m \cite{yfcc}& Unified video quality assessment\\
         LIVE-VQC \cite{vqc} &2018 &585 &videos of unique content, captured by users &Quality assessment of real world videos \\
         YouTube-UGC \cite{ytugccc} &2019 &1380 &Videos covering popular categories like Gaming, Sports&Quality assessment of user generated videos\\
         LSVQ \cite{pvq} &2021 &39075 &Real-world distorted videos and video patches &Quality assessment of user generated videos\\
         LIVE-NFLX-I \cite{bampis2017study} &2017 &112 &Videos with highly diverse and contemporary content & Quality-of-experience assessment of streaming\\
         LIVE-NFLX-II \cite{bampis2021towards} &2018 &588 &Videos with different bitrate and network conditions  & Quality-of-experience assessment of streaming\\
         WaterlooSQoE-III \cite{duanmu2018quality} &2018 &450  &Videos with diverse distortions, network conditions& Quality-of-experience assessment of streaming\\
         WaterlooSQoE-IV \cite{duanmu2020waterloo} &2019 &1,450 & Streaming videos generated from different setups& Quality-of-experience assessment of streaming\\
         TaoLive \cite{mdvqa} &2022 &3,762&Videos collected in real live streaming scenarios & Quality assessment of live streaming video\\
         OAVQAD \cite{zhu2023perceptual} &2023 &375& Distorted omnidirectional audio-visual sequences &Omnidirectional audio-visual quality assessment \\
         THQA \cite{thqa} & 2024 &800 &AI-generated talking head videos &Generative digital human visual quality assessment\\\hline
         \multicolumn{5}{l}{\textit{Quality assessment for \textbf{3D Contents}}}\\ \hdashline
         G-PCD \cite{alexiou2017performance,alexiou2017towards} &2017 &40 &Colorless point clouds& Colorless Point Cloud Quality Assessment\\
         RG-PCD \cite{alexiou2018point} &2018 &24 &Colorless point clouds& Colorless Point Cloud Quality Assessment\\
         VsenseVVDB \cite{zerman2019subjective} &2019 &32 &Compressed volumetric videos &Volumetric video quality assessment\\
         IRPC \cite{javaheri2020point} & 2020&108 &Point clouds under different rendering, coding solutions&Point cloud quality assessment\\
         WPC \cite{wpc} & 2021& 740& 3 different types of distorted point clouds &Colorful Point Cloud Quality Assessment\\
         WPC2.0 \cite{liu2021reduced} &2021 &400 &Video-based compressed point clouds  &Reduced-Reference Point cloud Quality Assessment\\
         SJTU-PCQA \cite{sjtupcqa} & 2020& 420& 6 different types of distorted point clouds &Colorful Point Cloud Quality Assessment\\
         LS-PCQA \cite{lspcqa} &2023 &1080 &31 different types of distorted point clouds & No-Reference Point cloud Quality Assessment\\
         BASICS \cite{basics} &2023 & 1494&Static point clouds& Point cloud Visual Quality Assessment  \\
         CMDM \cite{nehme2020visual} &2020&80 &Humans, Animals, Statues & Colored Mesh Quality Assessment\\
         TMQA \cite{nehme2023textured} & 2023&3,000 &Statues, Animals, Daily objects & Textured Mesh Quality Assessment \\
         DHH-QA \cite{dhhqa} &2023 &1540 & Scanned 3D human heads&3D Digital human quality assessment\\
         DDHQA \cite{ddhqa} &2023 &800 & Dynamic 3D digital human&3D Digital human quality assessment\\
         SJTU-H3D \cite{clip3dqa} &2023 &1120 & Static digital human&3D Digital human quality assessment\\
         6G-DHQA \cite{zhang2024quality} &2024 &400 & Digital twin dynamic characters&Quality of Service for 3D Streaming in 6G Network\\
         \bottomrule
    \end{tabular}}
    \label{tab:qa}
    \vspace{-15pt}
\end{table}

\subsection{Image Quality Assessment}
Image quality assessment (IQA) aims at predicting the technical visual quality of images, which consistently attracts considerable attention \cite{chahine2024deep}. Wang $et$ $al.$ \cite{wang2002image} discuss the limitations of error sensitivity-based frameworks in IQA. Despite these challenges, researchers continue to develop objective metrics to predict perceived quality. Sheikh $et$ $al.$ \cite{sheikh2006statistical} conduct a comprehensive subjective quality study with 779 distorted images to benchmark future research. Wang $et$ $al.$ \cite{wang2006modern} focus on computational models for predicting perceptual quality, detail leading algorithms, and propose new research directions. Moorthy $et$ $al.$ \cite{moorthy2009visual,moorthy2010two} introduce visual importance pooling strategies and construct blind IQA indices using natural scene statistics, evaluated on the LIVE database. Researchers explore various approaches to enhance IQA, including measuring image sharpness in the frequency domain \cite{de2013image}, utilizing relative gradient statistics and Adaboosting neural networks for blind IQA \cite{freund1997decision,liu2016blind}, and employing deep learning for distortion-generic blind IQA \cite{bianco2018use}. Mohammadi $et$ $al.$ \cite{mohammadi2014subjective} provide an overview of quality assessment methods for conventional and emerging technologies like HDR and 3D images. Zhai $et$ $al.$ \cite{zhai2020perceptual} emphasize the importance of perceptual quality assessment in visual communication systems.

Recent studies develop IQA methods focusing on more diverse image content, such as deep learning-based approaches for VR images \cite{kim2019deep} and blind 360-degree IQA using multi-channel convolutional neural networks \cite{sun2019mc360iqa}. Research on smartphone photography \cite{spaq} and face image quality \cite{schlett2022face}, including perceptual full-reference tasks \cite{cheon2021perceptual} and unsupervised techniques \cite{cheon2021perceptual}, remains significant. Advancements in no-reference IQA include multiscale feature representation \cite{cheon2021perceptual,zhang2021no}, multi-dimension attention networks \cite{cheon2021perceptual,lu2022deep,lu2021cnn,zhang2022nosuper}, and CNN \cite{cao2022no} via object detection. Overall, the literature on IQA highlights ongoing efforts to improve accuracy and reliability, employing both traditional and advanced techniques to address the associated challenges.

\subsection{Aesthetic Quality Assessment}
Research on aesthetic quality assessment traditionally focuses on general photo sets without specific content considerations. Li $et$ $al.$ \cite{li2010aesthetic} shift this focus to consumer photos with faces, while Jin $et$ $al.$ \cite{jin2010learning} develop methods for learning artistic portrait lighting templates, emphasizing Haar-like local lighting contrast features. Su $et$ $al.$ \cite{su2011scenic} propose an aesthetic modeling method for scenic photographs, covering both implicit and explicit features through a learning process.
In the realm of CNNs, Kao $et$ $al.$ \cite{kao2015visual} and Dong $et$ $al.$ \cite{dong2015photo} explore CNNs for aesthetic quality assessment. Kao $et$ $al.$ treat it as a regression problem \cite{kao2015visual}, and Dong $et$ $al.$ focus on understanding photo quality \cite{dong2015photo}. Lienhard $et$ $al.$ \cite{lienhard2015low} predict aesthetic quality scores of facial images by computing features related to technical aspects. Building on these foundations, Kao $et$ $al.$ propose a hierarchical CNN framework, demonstrating superior performance \cite{kao2016hierarchical}, and later incorporate semantic information into aesthetic assessments \cite{kao2017deep}.

Personalized algorithms for aesthetic assessment are introduced by Park $et$ $al.$ \cite{park2017personalized}, aiming to increase user satisfaction. Jin $et$ $al.$ predict aesthetic score distributions using CNNs for a nuanced approach \cite{jin2018predicting} and later introduce Aesthetic Attributes Assessment with the DPC-Captions dataset \cite{jin2019aesthetic}. Wu $et$ $al.$ develop a method using deep convolutional neural networks (DCNNs) for predicting product design awards \cite{wu2020product}. Kim $et$ $al.$ explore subjectivity in aesthetic quality assessment using a large-scale database \cite{kim2018objectivity}.
Kuang $et$ $al.$ present a deep multimodality learning approach for UAV video aesthetic assessment \cite{kuang2019deep}. Wang $et$ $al.$ propose a multidimensional aesthetic quality assessment method for mobile game images, focusing on fineness, color harmony, and overall quality \cite{wang2022deep}. Jin $et$ $al.$ introduce a new aesthetic mixed dataset and train a meta-reweighting network to address image aesthetic quality evaluation challenges \cite{jin2022pseudo}. Wu $et$ $al.$ introduce the PEAR framework, combining aesthetic rating and image reconstruction \cite{wu2022pear}. Chambe $et$ $al.$ highlight the use of deep learning in assessing professional photograph aesthetics using the AVA dataset \cite{chambe2022deep}. These studies underscore the diversity of approaches in aesthetic quality assessment across various domains.

\subsection{Video Quality Assessment}
Video quality assessment (VQA) is essential in diverse image and video processing applications, such as compression, communication, printing, analysis, registration, restoration, and enhancement \cite{li2024ntire,conde2024ais,zhang2022surveillance}. Wang $et$ $al.$ \cite{wang2004video} propose a new approach for designing VQA metrics, emphasizing structural distortion as a measure of perceived visual distortion. Seshadrinathan $et$ $al.$ \cite{seshadrinathan2007structural} introduce a quality metric for video sequences that incorporates motion information, highlighting the significance of motion in VQA.
Advancements in VQA methodologies increasingly integrate models of human visual perception. For instance, Wang $et$ $al.$ \cite{wang2007video} suggest using a statistical model of human visual speed perception in VQA frameworks. Zhai $et$ $al.$ \cite{zhai2008cross} expand VQA across multiple dimensions, assessing video quality across different spatial and temporal resolutions. Ninassi $et$ $al.$ \cite{ninassi2009considering} develop a perceptual full-reference VQA metric evaluating temporal distortions at eye fixation levels, emphasizing the importance of temporal variations in spatial visual distortions. Vu $et$ $al.$ \cite{vu2011spatiotemporal} introduce a spatiotemporal most-apparent-distortion model that considers motion artifacts to estimate motion-based distortion in videos.
Wu $et$ $al.$ \cite{wu2022fasterquality,wu2022fastquality,wu2023dover} employ fragment sampling to improve the efficiency of VQA models. Zhou $et$ $al.$ \cite{zhou2024light,dong2023light} carry out a VQA model for exposure correction evaluation, further enhancing the adaptability of VQA methods.

Subjective evaluations remain crucial in VQA algorithm assessment. Seshadrinathan $et$ $al.$ \cite{livevqa} conduct a study with human observers on distorted video sequences, leading to the creation of the LIVE Video Quality Database. Chikkerur $et$ $al.$ \cite{chikkerur2011objective} propose a classification scheme for objective VQA methods based on whether they consider natural visual or human visual system characteristics.
Recent advancements include display device-adapted video quality-of-experience assessments. Rehman $et$ $al.$ \cite{rehman2015display} introduce SSIMplus, a full-reference measure predicting real-time perceptual video quality based on human visual system behaviors, video content characteristics, display device properties, and viewing conditions. Bampis $et$ $al.$ \cite{bampis2017speed} develop reduced-reference models like SpEED-QA, which efficiently compute perceptually relevant quality features using local spatial operations on image frames and frame differences.

\subsection{3D Quality Assessment}
3D quality assessment (3DQA) has become increasingly popular with the advent of virtual reality (VR), augmented reality (AR), and the metaverse \cite{zhou2023perceptual}. Alexiou $et$ $al.$ \cite{alexiou2018pointa} observe a rising interest in point clouds, leading to the development of objective quality metrics. Su $et$ $al.$ \cite{su2019perceptual} advance this field by creating a comprehensive 3D point cloud database for subjective user studies, thereby facilitating future research. Diniz $et$ $al.$ \cite{diniz2020multi} introduce a quality metric based on multiple distances between reference and test point clouds, adapting the LBP descriptor for non-uniform point distributions.
Recent advancements in machine learning and deep learning have significantly impacted 3DQA. Lu $et$ $al.$ \cite{lu2020machine} propose an assessment method based on vision tasks to evaluate machine perception of point cloud quality. Liu $et$ $al.$ \cite{zhang2021ms} develop PQA-Net, a no-reference point cloud quality assessment model utilizing multi-view projection. Zhang $et$ $al.$ \cite{zhang2021ms} further develop MS-GraphSIM, a multiscale model that considers geometric and color features to accurately predict human perception. Additionally, Zhang $et$ $al.$ \cite{zhang2022no,zhang2021nomesh} create a no-reference metric for colored 3D models, including point clouds and meshes. Javaheri $et$ $al.$ \cite{zhang2021ms} propose a point-to-distribution metric that outperforms existing metrics.

To address domain adaptation challenges, Yang $et$ $al.$ \cite{yang2022no} present a no-reference assessment approach using natural images as the source domain and point clouds as the target domain. Zhang $et$ $al.$ explore different modalities to predict point cloud quality from rendered images or videos \cite{zhang2023evaluating,fan2022no,zhang2023simple,zhang2024optimizing}. Furthermore, Zhang $et$ $al.$ \cite{zhang2022mm} propose MM-PCQA, a multi-modal learning approach that combines 2D texture and semantic information with 3D geometry distortion sensitivity. More recently, attention is directed toward the efficiency of 3DQA with advancements like those proposed by Zhang $et$ $al.$ \cite{zhang2023eep,zhang2024gms}. Additional works focus on 3D digital human quality assessment \cite{zhang2024reduced,zhang2023geometry,chen2023no}.
Overall, the literature on 3DQA showcases a broad spectrum of approaches, from traditional metrics to advanced deep learning models, all aiming to accurately evaluate the quality of 3D models.

\begin{figure}
    \centering
    \includegraphics[width=.8\linewidth]{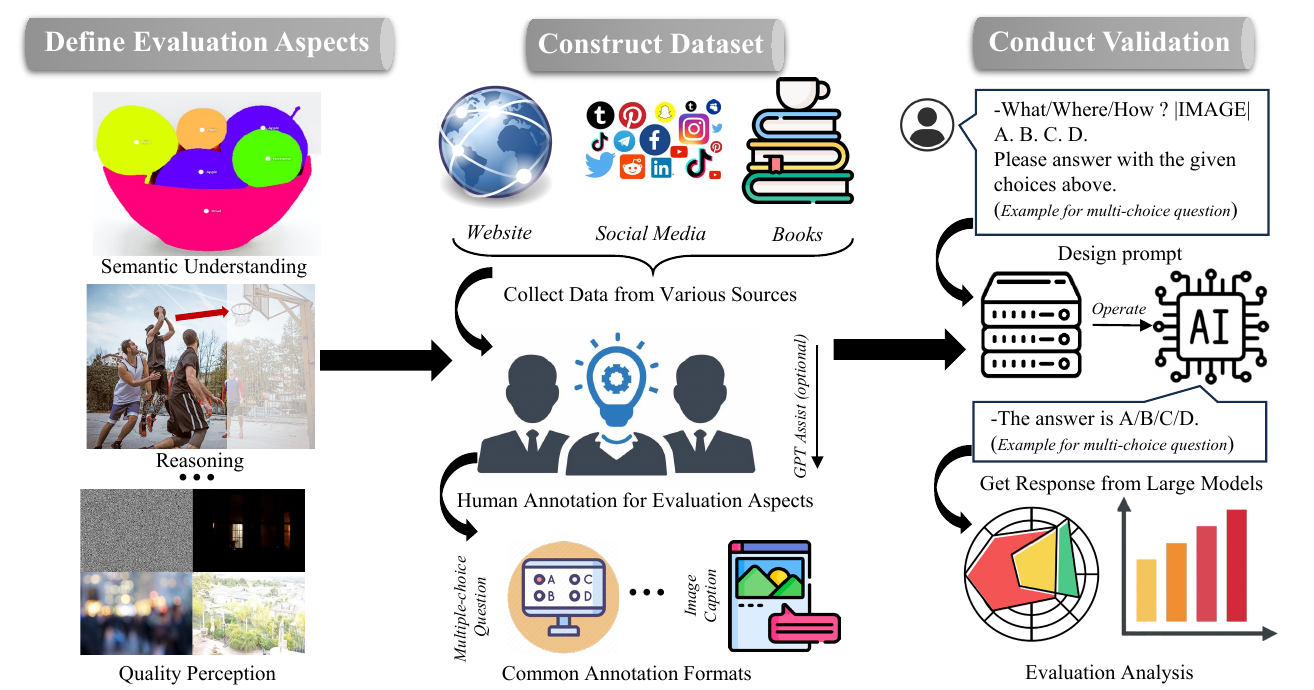}
    \caption{Illustration of the typical process of assessing large models.}
    \label{fig:benchmark}
    \vspace{-15pt}
\end{figure}

\section{Assessment of Large Models}
Assessment is crucial in the realm of large models, particularly for Large Multi-modal Models (LMMs) designed to tackle multi-modal challenges and engage with users. Evaluating LMMs is complex due to the diverse aspects of assessment and the lack of a standardized evaluation framework. Consequently, numerous multi-modal benchmarks have been introduced to assess both general and specific capabilities through various validation strategies. Despite these efforts, identifying an optimal solution remains an ongoing challenge. Additionally, the emergence of text-to-image/video and image-to-image generative large models has led to the creation of diverse visual AI-generated content (AIGC). Unlike traditional multimedia, visual AIGC content presents unique quality challenges, such as text alignment, generative-specific distortions, and unnaturalness. Addressing these issues necessitates the development of innovative quality assessment methods. In this section, we first provide an overview of the development of large multi-modal models. Next, we examine multi-modal benchmarks, with a particular focus on their ability-related aspects. Finally, we discuss the evaluation methods for visual AIGC.

\subsection{Large Muli-modal Models}
Large language models (LLMs) such as GPT-4~\cite{openai2023gpt4}, T5~\cite{flant5}, and LLaMA~\cite{llama} demonstrate remarkable linguistic proficiency across general human knowledge domains. These models extend their capabilities to multimodal tasks by incorporating visual inputs through CLIP~\cite{CLIP} and additional adaptation modules, as seen in large multimodal models (LMMs) \cite{otter,llamaadapterv2,llava,iblip,xcomposer}.
Specifically, OpenFlamingo \cite{openflamingo} integrates gated cross-attention dense blocks into the pretrained language encoder layers. InstructBLIP \cite{iblip} builds on BLIP-2 \cite{blip2} by adding vision-language instruction tuning. To advance \textit{open-source} LMMs, numerous projects utilize GPT-4 \cite{openai2023gpt4} for generating data to fine-tune vision-language models, exemplified by the LLaVA series \cite{llava,improvedllava,liu2024llavanext}. To effectively benchmark the diverse and rapidly developed LMMs, it is essential to employ robust assessment approaches. These methods should not only highlight the strengths and weaknesses of the models but also guide the direction for future improvements. Such evaluations are crucial for understanding and enhancing the capabilities of LMMs in various applications.

\subsection{Multi-modal Benchmarks}
A comprehensive comparison of mainstream benchmarks is presented in Table \ref{tab:bench}. These multimodal benchmarks facilitate the quality assessment of LMMs through a structured approach. The process begins with defining evaluation criteria, followed by the collection of relevant instances. Subsequently, either human annotators or other LLMs assist in annotating the data. This is followed by crafting specific prompts (\textit{multiple-choice questions, visual questions, multi-round dialogue}) designed to elicit responses from LMMs. Finally, a detailed analysis is conducted on the gathered responses to assess the effectiveness and accuracy of the LMMs.

\begin{enumerate}
    \item \textbf{Assessment for image understanding capabilities.} Image understanding traditionally includes tasks such as classification, detection, and identification. Early multi-modal benchmarks primarily address fundamental aspects of image comprehension, such as captioning significant elements and responding to simple queries \cite{cococaps,goyal2017making,nocaps}. However, these datasets mainly focus on extracting knowledge contained within the images themselves.
    To evaluate large models effectively, it becomes essential to test external knowledge. This includes recognizing well-known figures and comprehending specialized terms that necessitate a global understanding. The OK-VQA benchmark \cite{okvqa} is introduced to fill this gap by offering a variety of visual questions that rely on external knowledge and require intricate reasoning.
    With the rapid advancements in large models, particularly in interactive capabilities, new multi-modal benchmarks have been introduced. These benchmarks \cite{mplugowl,yu2023mm,shao2023tiny} challenge the models with open-ended questions that demand comprehensive, thoughtful responses to foster deeper discussions and analysis.
    Despite these developments, evaluating the responses of large models to open-ended questions remains a formidable challenge. OwlEval \cite{mplugowl} employs manual evaluation metrics to measure the quality of responses across various models. Additionally, some benchmarks utilize established NLP metrics like BLEU-4 \cite{papineni2002bleu} and CIDEr \cite{vedantam2015cider} to evaluate answers against set correct responses.
    The effectiveness of generative models, particularly GPT \cite{openai2023gpt4}, has been highlighted in recent research \cite{wu2023qbench,mmbench}, which shows improved evaluation outcomes that align more closely with human judgment.
    More recent benchmarks \cite{wu2023qbench,huang2024aesbench,zhang2024benchmark,zhang2024bench} have adopted multiple-choice formats. This shift offers advantages such as easier annotation and the straightforward transformation of visual questions and image captions into multiple-choice questions. Closed-answer formats enhance the accuracy of output assessments from large models and simplify the derivation of performance statistics based on question correctness.
    In addition, the focus of benchmarks has evolved from simple to complex, from general to specific, and from coarse-grained to fine-grained. This progression signifies a move towards more detailed and specialized evaluations of large models. 
    \item \textbf{Assessment for video understanding capabilities.} The multi-modal benchmarks for video understanding are still evolving. SEED-Bench \cite{seedbench} focuses on temporal understanding, encompassing action recognition, action prediction, and procedure understanding. MVBench \cite{li2024mvbench} extends these capabilities by evaluating the temporal comprehension of LMMs across 20 challenging video tasks, introducing a novel static-to-dynamic task generation method and utilizing automatic conversion of public video annotations into multiple-choice questions for fair and efficient assessment. Video-MME \cite{fu2024video} further explores LMM capabilities in video processing, offering a variety of video types across six visual domains and durations ranging from seconds to an hour. MLVU \cite{zhou2024mlvu} tackles the challenges in evaluating long video understanding by extending video lengths, incorporating diverse video genres such as movies and surveillance footage, and developing varied evaluation tasks to comprehensively assess LMMs' performance in long-video understanding.
    \item \textbf{Assessment for science-related capabilities.}  Multi-modal benchmarks designed to evaluate science-related capabilities require LMMs to accurately comprehend science content depicted in images and understand relevant scientific terminology. MathVista \cite{lu2023mathvista} and Math-Vision \cite{wang2024measuring} assess the math reasoning abilities of LMMs through a collection of high-quality mathematical problems set in visual contexts from real math competitions, which cover a range of mathematical disciplines and difficulty levels. More recently, MMMU \cite{mmmu} has been developed to include 11.5K meticulously curated multimodal questions derived from college exams, quizzes, and textbooks. These questions span six core disciplines: Art \& Design, Business, Science, Health \& Medicine, Humanities \& Social Science, and Technology \& Engineering, providing a comprehensive scope for evaluating the academic understanding capabilities of LMMs.
    \item \textbf{Assessment for hallucination-aware capabilities.} When investigating LMMs, researchers have noted a significant language bias, where the models' reliance on language priors often overshadows the visual context. To address this, RAVIE \cite{liu2023mitigating} and HallusionBench \cite{guan2024hallusionbench} have been introduced to perform quantitative analyses of the models' response tendencies, logical consistency, and various failure modes related to hallucinations. These benchmarks aim to assess and mitigate the imbalance between language and visual processing in LMMs.
    \item \textbf{Assessment for 3D understanding capabilities.} The area of 3D understanding is gaining traction to enhance the decision-making capabilities of autonomous agents. Yet, current 3D datasets and methodologies are often constrained to narrow applications. To broaden this scope, LAMM-Bench \cite{yin2024lamm} and M3DBench \cite{li2023m3dbench} have been introduced to evaluate large models' proficiency in interpreting multimodal 3D content, thereby paving the way for LMMs to act as generalists in a wider array of 3D tasks.
\end{enumerate}

\begin{table}[]\small
    \centering
    \renewcommand\tabcolsep{5pt}
    \renewcommand\arraystretch{1.1}
    \caption{Brief comparison of the multi-modal benchmarks. The scale is recorded as the size of the valid instances in the test set, if applicable.}
    \resizebox{\linewidth}{!}{\begin{tabular}{l|c|c|c|c}
         \toprule
         \textbf{Dataset} & \textbf{Year} & \textbf{Scale} & \textbf{Format} & \textbf{Evaluation Dimensions} \\ \hline
         \multicolumn{5}{l}{\textit{Assessment for \textbf{image understanding capabilities}}}\\ \hdashline
         COCO Caption  \cite{cococaps} & 2015 & 40K & Image captions & Image caption ability with common understanding\\
         VQA v2 \cite{goyal2017making} & 2017 & 453K & Visual questions & Visual question answering ability with common understanding\\
         Nocaps \cite{nocaps} & 2019 & 10.6K & Image captions & Image caption ability with common understanding\\
         OK-VQA \cite{okvqa} & 2019 & 14K & Visual questions &  Visual question answering ability with open knowledge\\
         OwlEval \cite{mplugowl} & 2023 & 82 & Open-ended questions & Multiple image understanding capabilities \\
         MME \cite{mme} & 2023 & 1,097& Yes-or-no questions & Perception and cognition abilities \\
         MMBench \cite{mmbench} & 2023 & 3,000 & Multiple-choice questions & Multiple bilingual multi-modal capabilities \\
         MM-Vet \cite{yu2023mm} & 2023 & 200 & Open-ended questions & Integrated capabilities for complicated tasks \\
         Tiny LVLM-eHub \cite{shao2023tiny} & 2023 & 2,100 & Open-ended questions & Multiple multi-modal capabilities \\
         DEMON \cite{li2023fine} & 2023 & 18.2K & Task instructions &  Demonstrative instruction understanding \\
         MagnifierBench \cite{li2023otterhd} & 2023 & 283 &Visual questions & Detecting minute details in high resolution images\\
         Q-Bench \cite{wu2023qbench} & 2023 & 3,489 & Multiple-choice questions & Low-level perception, description, and assessment abilities \\
         Aes-Bench \cite{huang2024aesbench} & 2023 & 2,800 & Multiple-choice questions & Image aesthetic perception and understanding\\
         BenchLMM \cite{cai2023benchlmm} & 2023 &  2,880 & Visual questions & Robustness against different image styles\\
          Q-Bench$^+$ \cite{zhang2024benchmark} & 2024 & 2,499 & Multiple-choice questions & Low-level perception comparison abilities \\
          MLLM-Bench \cite{ge2023mllm} & 2024 & 420 & Visual questions & Multiple multi-modal capabilities\\
          A-Bench \cite{zhang2024bench} & 2024 & 2,864 & Multiple-choice questions & Abilities of evaluating AI-generated Images \\
          GenAI-Bench \cite{lin2024evaluating} & 2024 &  1,600 & Visual questions & Visual-language alignment judging ability \\
          \hline
         \multicolumn{5}{l}{\textit{Assessment for \textbf{video understanding capabilities}}}\\ \hdashline
         SEED-Bench \cite{seedbench} & 2023 & 19K & Multiple-choice questions & Comprehension for both the image and video modality \\
         MVBench \cite{li2024mvbench} & 2023 & 4,000 & Multiple-choice questions & Multi-modal video understanding ability \\
         Video-MME \cite{fu2024video} & 2024 & 2,700 & Multiple-choice questions & Comprehensive video analysis abilities \\
         MLVU \cite{zhou2024mlvu} & 2024 & 2,593 & Mixed questions & Long video understanding capabilities \\ \hline
         \multicolumn{5}{l}{\textit{Assessment for \textbf{science-related capabilities}}}\\ \hdashline
         MathVista \cite{lu2023mathvista} & 2023 & 6,141 & Math problems & Mathematical reasoning abilities in visual contexts\\
         MMMU \cite{mmmu} & 2023 & 11.5K & Visual questions & College-level subject knowledge and deliberate reasoning abilities \\
          Math-Vision \cite{wang2024measuring} & 2024 & 3,040 & Math problems & Mathematical reasoning abilities in visual contexts \\ \hline
         \multicolumn{5}{l}{\textit{Assessment for \textbf{hallucination-aware capabilities}}}\\ \hdashline
         GRAVIE \cite{liu2023mitigating} & 2023 & 1,000 &  Open-ended questions &Hallucination and instruction following abilities \\
         HallusionBench \cite{guan2024hallusionbench} & 2024 & 1,129 & Visual questions & Image-context reasoning ability for hallucination \\ \hline
         \multicolumn{5}{l}{\textit{Assessment for \textbf{3D understanding capabilities}}}\\ \hdashline
         LAMM-Bench \cite{yin2024lamm} & 2023 & - & 2D/3D vision questions & 2D/3D vision abilities \\
         M3DBench \cite{li2023m3dbench} & 2023 & 1,500 & Task instructions & Multiple 3D understanding capabilities \\
         \bottomrule
    \end{tabular}}
    \label{tab:bench}
    \vspace{-15pt}
\end{table}

\begin{table}[]\small
    \centering
    \renewcommand\tabcolsep{8pt}
    \renewcommand\arraystretch{1.1}
    \caption{Brief comparison of quality assessment datasets for visual AIGC.}
    \resizebox{\linewidth}{!}{\begin{tabular}{l|c|c|c|c|c}
         \toprule
         \textbf{Dataset} & \textbf{Year} & \textbf{Scale} & \textbf{Ratings} & \textbf{Models} & \textbf{Quality Assessment Aspects} \\ \hline
         \multicolumn{5}{l}{\textit{Quality assessment for \textbf{AIGIs}}}\\ \hdashline
         DiffusionDB~\cite{wang2022diffusiondb} & 2022 & 14M & None & 1 & None\\
         HPS~\cite{wu2023human} & 2023 & 98.8K & 98.8K & 1 & Human preference \\
         ImageReward~\cite{xu2024imagereward} &2023& 136.9K & 136.9K & 3 & Human preference \\
         Pick-A-Pic~\cite{kirstain2024pick} & 2023 & 500K & 500K & 6 & Human preference \\
         AGIQA-1K~\cite{zhang2023perceptual}  &2023 & 1,080 & 23.7K & 2 & Overall perceptual quality\\
         AGIQA-3K~\cite{li2023agiqa} & 2023 & 2,982 & 125K & 6 & Perceptual quality and text alignment\\
         AIGCIQA2023~\cite{wang2023aigciqa2023} & 2023 & 2,400 & 48K & 6 & Perceptual quality, authenticity and correspondence \\
         AGIN~\cite{chen2023exploring} & 2023 & 6,049 & 181K & 18 & Overall naturalness \\
         AIGIQA-20K~\cite{AIGIQA-20K} & 2024 & 20K & 420K & 15 & Perceptual quality and text alignment\\
         AIGCOIQA2024~\cite{yang2024aigcoiqa2024} & 2024 & 300 & 6,000 & 5 & Perceptual quality, comfortability, and text alignment\\ 
         CMC-Bench~\cite{li2024cmc} & 2024 & 58K & 160K & 6 & Ultra-low bitrates compression quality \\ \hline
         \multicolumn{5}{l}{\textit{Quality assessment for \textbf{AIGVs}}}\\ \hdashline
         Chivileva $et$ $al.$~\cite {chivileva2023measuring} & 2023 & 1,005 & 48,2K & 5 & Perceptual quality and text alignment \\
         EvalCrafter~\cite{liu2023evalcrafter} & 2023 & 3,500 & 8,647 & 7 & Perceptual quality, text alignment, and temporal quality \\
         FETV~\cite{liu2024fetv} & 2023 & 2,476 & 29.7K & 3 & Perceptual quality, text alignment, and temporal quality \\
         VBench~\cite{huang2024vbench} & 2023 & 6,984 & - & 4 & Video quality and consistency \\
         T2VQA-DB~\cite{kou2024subjective} & 2024 & 10K & 540K & 9 & Peceprual quality and text alignment \\
         GAIA~\cite{chen2024gaia} & 2024 & 9,180 & 971K & 18 & Video action quality \\
         \bottomrule
    \end{tabular}}
    \label{tab:aigc_database}
    \vspace{-15pt}
\end{table}

\subsection{Visual AIGC Evaluation}
\textit{One look is worth a thousand words.} Drawing inspiration from this timeless proverb, numerous researchers have concentrated their efforts on developing text-to-image/video (T2I/V) models that vividly translate textual descriptions into visual representations.

\subsubsection{T2I/V Mdoels Development}

Notable innovations such as AlignDRAW \cite{mansimov2015generating} and text-conditional GAN \cite{reed2016generative} have pioneered unique architectural approaches to image synthesis. The field has seen further advancements with the introduction of stable diffusion models \cite{saharia2022photorealistic, rombach2022high}, significantly pushing the boundaries of T2I technology. On the commercial side, major companies are utilizing extensive datasets to develop and deploy highly effective T2I large models, including DALL-E \cite{gen:dalle}, Midjourney \cite{gen:MJ}, and Parti \cite{yu2022scaling}, among others. More recently, many efforts have been put into developing text-to-video large models. 
Extending pre-trained T2I models with temporal components is a standard method. CogVideo~\cite{hong2022cogvideo}, built on CogView2~\cite{ding2022cogview2}, introduces a multi-frame-rate hierarchical approach to better synchronize text-video sequences. Make-a-video~\cite{singer2022make} incorporates efficient spatial-temporal modules into a diffusion-based T2I architecture, specifically DALLE-2~\cite{ramesh2022hierarchical}. VideoFusion~\cite{luo2023videofusion} similarly utilizes DALLE-2 and implements a segmented diffusion method. A series of models including LVDM~\cite{he2022latent}, Text2Video-Zero~\cite{khachatryan2023text2video}, Tune-A-Video~\cite{wu2023tune}, AnimateDiff~\cite{guo2023animatediff}, Video LDM~\cite{blattmann2023align}, MagicVideo\cite{zhou2022magicvideo}, ModelScope~\cite{wang2023modelscope}, and VidRD~\cite{gu2023reuse}, draw from the advancements of stable diffusion \cite{rombach2022high} in video creation. Show-1~\cite{zhang2023show} merges pixel-based and latent-based approaches within video diffusion models. LaVie~\cite{wang2023lavie} modifies the core transformer block to accommodate spatial-temporal dynamics. Recently, OpenAI introduced Sora~\cite{videoworldsimulators2024}, an impressive T2V model capable of producing 60-second high-fidelity videos, setting a new direction in T2V technology.

\subsubsection{Quality assessment datasets for visual AIGC} 
To tackle the challenge of quality assessment for visual AIGC, many quality assessment datasets have been proposed during the last two years. A brief illustration of these datasets is shown in Table~\ref{tab:aigc_database}.
\begin{enumerate}
    \item \textbf{Quality assessment datasets for AI-generated images (AIGI):}
In recent years, multiple AIGI datasets have been introduced. The DiffusionDB~\cite{wang2022diffusiondb} dataset is launched as the inaugural large-scale text-to-image prompt dataset, comprising 14 million AIGIs created by stable diffusion based on real user prompts and hyperparameters. The HPS~\cite{wu2023human} dataset gathers 98,807 AIGIs generated in the Stable Foundation Discord channel, accompanied by 25,205 selections made by humans. ImageReward~\cite{xu2024imagereward} offers a dataset with 137k prompt-image pairings derived from DiffusionDB, where each pair is evaluated on three criteria: overall rating, image-text alignment, and fidelity. The Pick-A-Pic~\cite{kirstain2024pick} dataset features over 500,000 instances and 35,000 unique prompts, with each instance comprising a prompt, two generated AIGIs, and a preference label. The AGIQA-1K~\cite{zhang2023perceptual}, AGIQA-3K~\cite{li2023agiqa}, and AIGCIQA2023~\cite{wang2023aigciqa2023} datasets include 1,080, 2,982, and 2,400 AIGIs respectively. Specifically, the AGIQA-1K dataset first proposes the three most important evaluation dimensions for AIGIs: technical quality, aesthetic quality, and text alignment. The AGIN~\cite{chen2023exploring} dataset assembles 6,049 AIGIs and performs an extensive subjective study to ascertain human perspectives on overall naturalness. More recently, the AIGIQA-20K~\cite{AIGIQA-20K} dataset has been released, featuring 20,000 AIGIs generated by 15 prominent T2I models, with MOSs obtained from 21 evaluators. To tackle the challenge of omnidirectional image quality assessment in VR/AR environment, the AIGCOIQA2024~\cite{yang2024aigcoiqa2024} dataset conducts comprehensive benchmarks by generating 300 images from 5 AIGC models using 25 text prompts, assessing human visual preferences in terms of quality, comfortability, and correspondence, and evaluating state-of-the-art IQA models' performance on this database. CMC-Bench~\cite{li2024cmc} proposes to evaluate the cooperative performance of Image-to-Text and Text-to-Image models for ultra-low bitrate image compression, demonstrating that some model combinations outperform advanced visual codecs and highlighting areas for further optimization in LMMs.
   \item \textbf{Quality assessment datasets for AI-generated videos (AIGV):}
In contrast to AIGI datasets, there are fewer AIGV datasets. Chivileva $et$ $al.$~\cite {chivileva2023measuring} introduce a dataset comprising 1,005 videos generated by five T2V models, with 24 users participating in a subjective study. EvalCrafter~\cite{liu2023evalcrafter} develops a dataset for user study from 500 prompts using five T2V models, resulting in a total of 2,500 AIGVs. Similarly, the FETV~\cite{liu2024fetv} dataset employs 619 prompts and four T2V models, with three users for annotation. VBench~\cite{huang2024vbench} is more extensive, encompassing approximately 1,700 prompts and four T2V models. To expand the scale, the T2VQA-DB~\cite{kou2024subjective} dataset has been proposed, featuring 10,000 AIGVs produced by nine different T2V models, with 27 subjects engaged in collecting the MOSs. More recently, the GAIA~\cite{chen2024gaia} dataset has been carried out to focus on the action quality of the generated videos. It's worth mentioning the first competition track on AIGI/AIGV quality assessment has been held by \textit{NTIRE 2024 Quality Assessment for AI-Generated Content - Track 1/2: Image/Video} \cite{liu2024ntire}.
\end{enumerate}
 
\subsubsection{Quality assessment methods for visual AIGC} 
To be candid, the development of quality assessment methods for visual AIGC has lagged significantly behind the needs of AIGC technology. Researchers continue to struggle with accurately evaluating the quality of visual AIGC outputs. Initially, popular metrics for T2I/V generation such as Inception Score (IS) \cite{salimans2016improved}, Frechet Video Distance (FVD) \cite{unterthiner2018towards}, and CLIP Similarity (CLIPSim) \cite{wu2021godiva}, are found inadequate in reflecting actual user preferences. IS, which employs the Inception Network \cite{szegedy2016rethinking}, generates a distribution that intends to capture image/video quality and diversity but has been criticized for its imprecision. FVD measures the similarity between the feature distributions of generated videos and natural videos using I3D features~\cite{carreira2017quo}, with a lower FVD indicating a more natural-looking video. However, obtaining a suitable natural video for comparison is often impractical. CLIPSim leverages the CLIP model~\cite{radford2021learning} to assess the alignment between the original text and the generated video content but falls short by neglecting temporal information and overall perceptual video quality.

To address these shortcomings, many researchers have reverted to traditional IQA methods. These traditional approaches, which evaluate technical distortions such as noise, blur, and semantic content, have provided partial solutions for AIGC evaluation in terms of technical quality. Recognizing the unique challenges of visual AIGC content, several IQA models have been specifically developed for AIGIs. Notably, HPS~\cite{wu2023human} and PickScore~\cite{kirstain2024pick} use CLIP-based~\cite{radford2021clip} models to mimic human preferences for generated images. ImageReward~\cite{xu2024imagereward} employs a BLIP-based~\cite{li2022blip} architecture to predict image quality. Furthermore, the capability of LMMs is being explored to enhance AIGC evaluation tasks. Q-bench~\cite{wu2023qbench} is the first to investigate LMMs' performance in assessing visual quality. Subsequent studies such as those by Wu $et$ $al.$~\cite{wu2023qinstruct,zhang2023qboost, wu2023qalign, wu2024towards} have introduced training procedures to utilize LMMs for IQA tasks. Additionally, Q-Refine~\cite{li2024qrefine} is a quality-enhanced refiner designed to guide the refining processes in T2I models. Wang $et$ $al.$~\cite{wang2024large} have further employed the coherence and semantic content discernment capabilities of LMMs to aid in the evaluation process. The details of these models are discussed in Section~\ref{sec:to_asses}.


Similarly, for the AIGV evaluation aspects, the traditional VQA methods are used to predict the spatial and temporal quality of the generated AIGVs. Although these methods are capable of solving the specific dimension, they still can not satisfy the AIGV evaluation since the evaluation is conducted comprehensively from many aspects.
Then there are several works targeting the VQA tasks of AIGVs. VBench~\cite{huang2024vbench} and EvalCrafter~\cite{liu2023evalcrafter} build benchmarks for AIGVs by designing multi-dimensional metrics. MaxVQA~\cite{wu2023maxvqa} and FETV~\cite{liu2024fetv} propose separate metrics for the assessment of video-text alignment and video fidelity, while T2VQA~\cite{kou2024subjective} handles the features from the two dimensions as a whole.  We believe the development of the VQA model for AIGV will certainly benefit the generation of high-quality videos.

\section{Large Models in Assessment Roles}
\label{sec:to_asses}

Considering the impressive capabilities of large models in visual understanding, it is logical to leverage them as evaluators for both traditional IQA/VQA tasks and emerging AIGC evaluation tasks. The benefits are clear: large models can interact with users and respond to quality-related inquiries, enhancing the scope and flexibility of assessments. However, a significant challenge arises because these models primarily produce textual outputs, whereas many specific tasks require quantifiable scores. The textual responses generated by these models can be ambiguous and lack the precision needed for definitive evaluations. Researchers have developed numerous innovative methods to employ large models as evaluators, which can broadly be categorized into two main strategies: 1) Prompt-driven Evaluation: This involves designing specific prompts that encourage large models to directly generate desired outcomes. By carefully crafting the input prompts, researchers can guide the models to produce outputs that align closely with evaluation objectives. 2) Feature-based Assessment: This strategy entails using large models to extract features that are indicative of quality. Once these quality-aware features are obtained, they can be used to perform regression analyses to quantitatively assess the quality of the content. These strategies showcase the flexibility and potential of large models in automating and enhancing content evaluation processes.

\subsection{Prompt-driven Evaluation}
Considering the various prompt techniques developed in the natural language processing (NLP) field, such as in-context learning~\cite{dong2022survey}, standard prompting~\cite{shin2022effect}, and chain-of-thought reasoning~\cite{wei2022chain,wang2022self}, it is common to adapt these methods to engage large models in quality assessment tasks.

\subsubsection{Single-stimulus Prompt-driven Evaluation} 
When evaluating single images, the prompt used for large models can be direct and straightforward. Q-Bench~\cite{wu2023qbench} offers a simple example of how to craft prompts for articulating the quality attributes of single images, as demonstrated in the sample provided below: 

\textit{-User: Assume that you are an expert in quality assessment. Please describe the quality, aesthetics, and other low-level appearance of the image |<IMAGE>| in detail. Then give the final quality rating based on your previous description. }

\textit{-Response: [The content of quality description] + [Quality Rating].}

\noindent This prompt endows large models with a specific role and divides the quality assessment into two parts: quality description and final rating. The quality description enables large models to thoroughly examine the images, forming the foundation for the subsequent rating. X-iqe~\cite{chen2023x} further utilizes the chain-of-thought approach by breaking down the quality prompt into multiple rounds of quality-aware dialogue. This sequence encompasses fidelity evaluation$\rightarrow$alignment evaluation$\rightarrow$aesthetic evaluation and follows the prompt order from image description$\rightarrow$task-specific analysis$\rightarrow$scoring. Here we present the sample chain-of-thought prompt for multi-dialogue quality assessment:

\textit{-User: Assume that you are an expert in quality assessment. Please describe the quality, aesthetics, and other low-level appearance of the image |<IMAGE>| in detail. }

\textit{-Response: [The content of quality description].}

\textit{-User: Based on your description, how is the (specific-quality-dimension) of this image? }

\textit{-Response: [Fine-grained analysis of (specific-quality-dimension)].}

\textit{-User: Give the rating of (specific-quality-dimension) and overall perceptual quality of the image.}

\textit{-Response: [Quality Rating for (specific-quality-dimension) and overall perceptual quality].}

\noindent The single-stimulus evaluation is simple and scalable, directly showcasing the quality understanding of large models, and can be easily adapted to many application scenarios. Promisingly, the single-stimulus prompt-driven evaluation is considered a leading candidate for future explainable no-reference quality assessment tasks.

\subsubsection{Multiple-stimulus Prompt-driven Evaluation}
In some scenarios, pairwise or listwise comparisons are essential for determining superior image quality~\cite{ye2014active,zhang2021full,zhang2023perceptual}. To address this, several researchers have developed specialized prompt systems for multiple-stimulus, prompt-driven evaluations. For example, 2AFC-LMMs~\cite{zhu20242afc} assess the IQA capabilities of LMMs using a two-alternative forced choice method. This approach involves devising coarse-to-fine pairing rules and employing maximum a posteriori (MAP) estimation~\cite{tsukida2011analyze} to convert pairwise preferences of different LMMs into global ranking scores. The pairwise comparison sample is illustrated as follows:

\textit{-User: Assume that you are an expert in quality assessment.}

\textit{\quad \quad \quad This is the first image: |<IMAGE1>|.}

\textit{\quad \quad \quad This is the second image: |<IMAGE2>|.}

\textit{\quad \quad \quad Which image has better visual quality?}

\textit{-Response: The first/second image.}

\noindent It's worth noting that in the prompt sample provided, it is crucial to specify which image is the first and which is the second. Without this clarification, LMMs may not be able to distinguish between the first and second images. Additionally, Wu $et$ $al.$ utilize MAP estimation under Thurstone’s Case V assumption~\cite{thurstone2017law} to efficiently aggregate pairwise rankings and propose a novel method to gather partial rankings by presenting a set of images to LMMs for simultaneous ranking.                                                                                                                \begin{table*}
    \centering
    \renewcommand\arraystretch{1.2}
    \renewcommand\tabcolsep{4.8pt}
    \caption{Performance of LMMs under the zero-shot setting on traditional quality assessment datasets, in comparison with NIQE and CLIP-ViT-Large-14 (the visual backbone of most LMMs). All the LMM models are tested with the Q-Bench~\cite{wu2023qbench} softmax-based strategy. Metrics are \textit{SRCC/PLCC}. Best in bold. }
    \vspace{-5pt}
    \resizebox{\linewidth}{!}{\begin{tabular}{l|cccc|cc|c|c}
    \toprule
    {\textbf{Dataset Type}}  & \multicolumn{4}{c|}{{In-the-wild}} & \multicolumn{2}{c|}{{Generated}} & \multicolumn{1}{c|}{{Authentic}} & \multirow{2}{27pt}{\textit{Average}}\\ \cdashline{1-8}
     \textbf{Model / Dataset}  &{\textit{KONiQ-10k}} & {\textit{SPAQ}} & {\textit{LIVE-FB}} & \textit{LIVE-itw} & {\textit{CGIQA-6K}} & {\textit{AGIQA-3K}} & {\textit{KADID-10K}} & \\ \hline 
    NIQE~\cite{niqe} & 0.32/0.38 & {0.69}/{0.67} & 0.21/0.29 & {0.48}/0.45 & 0.08/0.06 & 0.56/0.52 & 0.37/0.43 & 0.39/0.40\\
    CLIP-ViT-Large-14~\cite{radford2021clip} & {0.47}/{0.51} & 0.39/0.39 & 0.22/0.24 & 0.31/0.31 & {0.29}/{0.29} & 0.44/0.46 & 0.38/0.39 & 0.35/0.37\\ \cdashline{1-9}
    InfiMM (\textit{Zephyr-7B})~\cite{InfiMM} & {0.51}/{0.55} & 0.62/0.63 & 0.27/0.30 & {0.55}/{0.58} & 0.23/0.25 & {0.71}/{0.77} & 0.47/0.45 & {0.48}/0.50\\
    Emu2-Chat (\textit{LLaMA-33B})~\cite{emu2} & \textbf{0.66}/\textbf{0.71} & {0.71}/{0.70} & {0.36}/{0.34} & {0.60}/{0.61} & 0.22/0.27 & \textbf{0.76}/{0.75} & \textbf{0.84}/\textbf{0.79} & \textbf{0.59}/\textbf{0.60}\\
    Fuyu-8B (\textit{Persimmon-8B})~\cite{fuyu-8b} & 0.12/0.12 & 0.13/0.18 & 0.16/0.13 & 0.23/0.18 & 0.12/0.12 & 0.37/0.32 & 0.10/0.09 & 0.17/0.16 \\
    BakLLava (\textit{Mistral-7B})~\cite{bakllava} & 0.39/0.39 & 0.41/0.40 & 0.23/0.22 & 0.34/0.34 & 0.18/0.21 & 0.54/0.56 & 0.34/0.36 & 0.35/0.35\\
    mPLUG-Owl2 \textit{(LLaMA-7B)}~\cite{mplug2} & 0.20/0.25 & 0.59/0.61 & 0.22/0.29 & 0.29/0.34 & -0.02/-0.03 & 0.47/0.49  &  0.54/0.55 & 0.33/0.36\\
    LLaVA-v1.5 (\textit{Vicuna-v1.5-7B)}~\cite{improvedllava} & {0.46}/0.46 & 0.44/0.47  & {0.31}/0.32 & 0.34/0.36 & \textbf{0.32}/\textbf{0.33} & {0.67}/{0.74} & 0.42/0.44 & 0.42/0.45\\
    LLaVA-v1.5 (\textit{Vicuna-v1.5-13B)}~\cite{improvedllava} & 0.45/{0.46} & 0.56/0.58  & {0.31}/{0.34} & 0.45/0.48 & {0.29}/{0.30} & 0.66/{0.75} & 0.39/0.40 & 0.44/0.47\\
    InternLM-VL \textit{(InternLM)}~\cite{xcomposer}  & {0.56}/{0.62} & \textbf{0.73}/\textbf{0.75} & \textbf{0.36}/\textbf{0.42} & \textbf{0.61}/\textbf{0.68} & 0.24/0.27 & {0.73}/\textbf{0.78} & {0.55}/{0.57} & {0.54}/{0.58}\\
    IDEFICS-Instruct \textit{(LLaMA-7B)}~\cite{idefics} & 0.38/0.40 & 0.47/0.48 & 0.24/0.24 & 0.41/0.43 & 0.24/0.23 & 0.56/0.62 & 0.37/0.37 & 0.38/0.40\\
    Qwen-VL \textit{(QwenLM)}~\cite{Qwen-VL}  & 0.47/0.55 & {0.68}/{0.67} & {0.30}/{0.34} & {0.50}/{0.53} & 0.27/0.28 & 0.62/0.69 & {0.49}/{0.49} & {0.48}/{0.51}\\
    Shikra (\textit{Vicuna-7B)}~\cite{shikra} & 0.31/0.31 & 0.32/0.34 & 0.24/0.24 & 0.32/0.34 & 0.20/0.20 & 0.64/0.66 & 0.32/0.33 & 0.34/0.35\\
    Otter-v1 \textit{(MPT-7B)}~\cite{otter} & 0.41/0.41 & 0.44/0.44 & 0.14/0.14 & -0.01/0.02 & 0.25/0.26 & 0.48/0.48 & {0.56}/{0.58} & 0.32/0.33\\
    Kosmos-2~\cite{kosmos2} & 0.26/0.28 & {0.64}/0.64 & 0.20/0.20 & 0.36/0.37 & 0.21/0.23 & 0.49/0.49 & 0.36/0.37 & 0.36/0.37\\
    InstructBLIP \textit{(Flan-T5-XL)}~\cite{iblip} & 0.33/0.36 & 0.58/0.60 & 0.25/0.27 & 0.11/0.11 & 0.17/0.19 & 0.38/0.40 & 0.21/0.18 & 0.29/0.30\\
    InstructBLIP \textit{(Vicuna-7B)}~\cite{iblip} & 0.36/0.44 & {0.68}/{0.69} & 0.20/0.28 & 0.25/0.37 & 0.26/{0.30} & 0.63/0.66 & 0.34/0.38 & 0.39/0.45\\
    VisualGLM-6B \textit{(GLM-6B)}~\cite{glm}  & 0.25/0.23 & {0.50}/{0.51} & 0.15/0.15 & 0.11/0.12 & 0.21/0.18 & 0.34/0.35 & 0.13/0.13 & 0.24/0.24\\
    mPLUG-Owl \textit{(LLaMA-7B)}~\cite{mplugowl} & 0.41/0.43 & 0.63/{0.64} & 0.24/0.27 & 0.44/{0.49} & 0.15/0.18 & {0.69}/{0.71} & 0.47/{0.49} & 0.43/0.46\\
    LLaMA-Adapter-V2~\cite{llamaadapterv2} & 0.35/0.36 & 0.46/0.51 & 0.28/{0.33} & 0.30/0.36 & 0.26/0.27 & 0.60/0.67 & 0.41/0.43 & 0.38/0.42\\
    LLaVA-v1 \textit{(Vicuna-13B)}~\cite{llava} & 0.46/0.46 & 0.44/0.46 & 0.26/0.28 & 0.40/0.42 & 0.21/0.24 & 0.63/0.68 & 0.35/0.37 & 0.39/0.42\\
    MiniGPT-4 (\textit{Vicuna-13B)}~\cite{minigpt4} & 0.24/0.26 & 0.24/0.25 & 0.17/0.18 & 0.34/0.34 & 0.25/0.25 & 0.57/0.59 & 0.24/0.23 & 0.29/0.30\\
    \bottomrule
    \end{tabular}}
    \vspace{-15pt}
    \label{tab:performance}
\end{table*}

\subsection{Feature-based Assessment}
\subsubsection{Assessment by CLIP} CLIP~\cite{radford2021clip} represents an early breakthrough in large models that align image and text modalities, though it lacks advanced reasoning and deep understanding. However, it is capable of aligning images with corresponding quality levels, which has spurred numerous quality assessment initiatives using CLIP and other pretrained vision-language models. ZEN-IQA~\cite{miyata2024zen} utilizes carefully constructed prompt pairs and triplets, enhancing the intuitiveness and interpretability of the evaluation process. QA-CLIP~\cite{pan2023quality} introduces a fine-grained quality-level stratification strategy and a two-stage training model, optimizing both text and image encoders to improve the accuracy of quality assessments across a wide range of levels. Miyata $et$ $al.$~\cite{miyata2023interpretable} employ a pre-trained vision-language model with a prompt pairing strategy using multiple antonym-prompt pairs to accurately estimate and elucidate the perceptual quality of images.
LIQE~\cite{liqe} capitalizes on auxiliary knowledge from scene classification and distortion type identification through vision-language correspondence, automatically optimizing model parameter sharing and loss weighting to bolster performance across various IQA datasets. PromptIQA~\cite{chen2024promptiqa} adapts to diverse IQA requirements without fine-tuning by leveraging a sequence of Image-Score Pairs (ISP) as prompts for targeted predictions and training on a mixed dataset with innovative data augmentation strategies.
Pan $et$ $al.$\cite{pan2024multi} improve semantic analysis by deploying a dual-prompt scheme and a multi-layer prompt structure in the visual branch to enhance adaptability and performance in image quality assessment, achieving robust and accurate results across diverse datasets. Wang $et$ $al.$\cite{wang2024large} augment traditional deep neural network-based IQA models with semantically informed guidance and a mixture of experts structure to dynamically integrate semantic and quality-aware features, significantly enhancing the assessment of AI-Generated images with superior generalization capabilities. Finally, QualiCLIP~\cite{agnolucci2024quality} utilizes a CLIP-based self-supervised, opinion-unaware method that employs a quality-aware image-text alignment strategy, training on synthetically degraded images aligned with quality-related antonym text prompts to produce representations that correlate with image quality.

Apart from IQA, vision-language models have been effectively utilized in diverse tasks such as IAA, VQA, and 3DQA. VILA~\cite{ke2023vila} leverages pre-trained image-text encoder-decoder models with image-comment pairs and a lightweight rank-based adapter for efficient adaptation, achieving powerful zero-shot capabilities in aesthetic tasks by learning from user comments instead of human-labeled scores. AesCLIP~\cite{sheng2023aesclip} enhances IAA by using a multi-attribute contrastive learning framework based on CLIP, which classifies aesthetic comments into attribute categories and learns attribute-aware representations to address the domain shift from general visual perceptions to specific aesthetic criteria. Hou $et$ $al.$~\cite{clipiaa} introduce a transparent deep learning framework for IAA that develops Tag-based Content Descriptors (TCDs), utilizing human-readable tags to explicitly describe image content, thus enhancing model interpretability and significantly improving assessment accuracy.
BVQI~\cite{bvqiplus,wu2023bvqi} and its localized version, BVQI-Local, employ the Semantic Affinity Quality Index (SAQI) methodology using CLIP to assess video quality by evaluating semantic content affinity with textual prompts, significantly outperforming traditional zero-shot indices by integrating low-level metrics and optimizing fine-tuning schemes for enhanced generalization. DHQA~\cite{clip3dqa} predicts the quality levels of digital humans by leveraging semantic and distortion features extracted from projections and geometry features from mesh structures, with the assistance of NIQE and CLIP. Zhang $et$ $al.$~\cite{zhang2024quality} propose the use of pre-trained vision-language models in the transmission systems of digital twins under a 6G environment, showcasing the broad applicability and potential of these models in cutting-edge technology scenarios.

\subsubsection{Zero-shot Assessment by LMMs} The Q-series~\cite{wu2023qbench,zhang2023qboost} introduces an innovative approach to employing LMMs as evaluators under zero-shot setting, presenting significant advancements that merit a comprehensive examination. Specifically, Q-Bench~\cite{wu2023qbench} identifies a key challenge: LMMs often struggle to produce sufficiently quantifiable outputs, whether instructed to provide direct textual ratings or numerical values. To address this issue, Q-Bench innovates by extracting the softmax-pooled results from the logits of the two most frequently predicted tokens (\textit{good} and \textit{poor}) as outlined in the model's response template. This approach not only aligns more closely with human judgment but also provides a more quantifiable measure of quality compared to the direct token outputs typically generated by LMMs.
The softmax-based strategy developed by Q-Bench has proven to be highly effective, better correlating with human perceptions and effectively bridging the gap between emerging LMMs and traditional IQA tasks. This method also serves as a robust benchmark for evaluating the zero-shot performance of LMMs in quality assessment tasks. 

By highlighting the performance of LMMs in Table~\ref{tab:performance} under this new strategy, we can swiftly gauge and enhance the quality assessment capabilities of current models. Primarily, it is observed that the majority of LMMs significantly outperform NIQE in non-natural circumstances (CGI, AIGC, artificial distortions), demonstrating their potential as general-purpose evaluators across a broader range of low-level appearances. Additionally, even without explicit alignment with human opinions during training, the top-performing LMM outperforms CLIP-ViT-Large-14 by a substantial margin (25\%). These results indicate that, although most MLLMs still utilize CLIP as visual encoders, their robust capabilities in language decoding significantly enhance their performance in visual quality assessment, even without specific training.
Q-Boost~\cite{zhang2023qboost} further expand the token set from \{\textit{good} $\leftrightarrow$ \textit{poor}\} (\textit{positive} $\leftrightarrow$ \textit{neutral} $\leftrightarrow$ \textit{negative}) to a more fine-grained token set \{\textit{good} + \textit{high} + \textit{fine} $\leftrightarrow$ \textit{average} + \textit{medium} + \textit{acceptable} $\leftrightarrow$ \textit{poor} + \textit{low} + \textit{bad}\} (\textit{positive} $\leftrightarrow$ \textit{neutral} $\leftrightarrow$ \textit{negative}). By incorporating the neural tone and expanding the synonym words, Q-Boost significantly improves the performance of LMMs on zero-shot quality assessment tasks.

\subsubsection{Quality-Infused LMM Assessment} Relying only on the inherent knowledge within LMMs for quality assessment tasks has proven inadequate. In response, researchers have developed several innovative strategies to infuse quality-aware knowledge into LMMs, thereby achieving state-of-the-art performance in quality assessment. Q-Instruct~\cite{wu2023qinstruct} enhances the perceptual capabilities of LMMs by introducing a large-scale, description-rich dataset. Q-Align~\cite{wu2023qalign} collects a multi-modal dataset to facilitate quality alignment, successfully reformatting traditional quality assessment datasets into text-labeled ones, thus fine-tuning LMMs to become effective evaluators with impressive performance and generalization capabilities. Co-instruct~\cite{wu2024towards} amasses a large-scale multi-modal dataset annotated by GPT-4v, supporting the pretraining of LMMs in comparative settings such as pairwise choices and listwise rankings, enabling LMMs to handle open-ended quality comparison questions and provide necessary rationale. DepictQA~\cite{you2023depicting} leverages LMMs to deliver detailed, language-driven evaluations that outperform score-based methods, closely mirroring human reasoning through a hierarchical task framework and diverse training data sources. DepictQA-Wild~\cite{you2024descriptive} addresses the shortcomings of existing methods by covering a broader range of IQA tasks, enhancing dataset quality, maintaining image resolution, and estimating confidence scores. Visualcritic~\cite{huang2024visualcritic} is engineered to reflect human-like perception of visual quality, offering both quantitative Mean Opinion Score measurements and qualitative evaluations with detailed explanations. Aesexpert~\cite{huang2024aesexpert} constructs expert-level aesthetic foundational models by assembling a rich corpus of aesthetic critique datasets, aligning LMMs with human aesthetic perceptions. UNIAA~\cite{zhou2024uniaa} introduces a cost-effective approach to transform existing datasets into unified, high-quality visual instruction tuning data, developing an expert-level aesthetics LMM. LMM-PCQA~\cite{zhang2024lmm} extends the application of LMMs to point cloud quality assessment (PCQA) by converting PCQA datasets into question-answer instructional pairs and deducing point cloud quality from cube-like projections. Compare2Score~\cite{zhu2024adaptive} utilizes large multimodal training and an innovative soft comparison method to convert relative image quality comparisons into a continuous, precise quality score, significantly surpassing existing models across various datasets. More recently, Sun $et$ $al.$~\cite{sun2024enhancing,sun2024dual} utilize the feature map from the last hidden layer of large models to tackle the challenge of both no-reference video quality assessment and portrait image quality assessment.

\section{Conclusion and Future Outlook}
In this survey, we explore the evolution and current state of quality assessment in the era of large models, emphasizing the critical roles of large multimodal models (LMMs) and large language models (LLMs). These models have significantly advanced our ability to assess and enhance the quality of multimedia content across various dimensions such as images, videos, and 3D content. Our review meticulously details several key points: 1) the shift from traditional quality assessment methods to advanced techniques that leverage large models, 2) the nuanced evaluation of these large models, and 3) the utilization of large models as evaluators themselves. Through an in-depth examination of multimodal benchmarks, we identify both the strengths and limitations of current models and outline the essential capabilities that these models need to develop further. The integration of LMMs has not only improved the precision and efficiency of quality assessments but also provided a closer alignment with human judgment, enabling the handling of complex, multimodal tasks that were previously challenging.

Looking forward, the field of quality assessment is poised for transformative advancements with the potential integration of even more sophisticated AI technologies. To summarize the future key points of quality assessment, we give the anticipated developments as follows:
\begin{enumerate}
    \item Refined Benchmarks: We advocate for an evolution of benchmarks from broad-based approaches to more detailed and streamlined standards. It is crucial to collaboratively establish a universally accepted benchmarking standard to address the challenges posed by the rapid advancement of large models effectively.
    \item Enhanced Multimodal Integration: Future advancements are likely to concentrate on deeper and more seamless integration of various modalities—text, image, video, and audio—to enable a more holistic and context-aware assessment of quality.
    \item Ethical Considerations and Bias Mitigation: Given the inherent subjectivity in quality assessment, increasing focus on ethical considerations and bias mitigation is essential. Ensuring that AI-driven quality assessment tools are developed and used ethically will become increasingly important, which includes addressing potential biases in AI models and the training data used for these systems.
\end{enumerate}
These directions highlight the necessity for continuous innovation and vigilance as we further integrate AI into quality assessment processes.


\bibliographystyle{ACM-Reference-Format}
\bibliography{sample-base}










\end{document}